# A new exactly integrable hypergeometric potential for the Schrödinger equation

T.A. Ishkhanyan[1,2], V.A. Manukyan[1], A.H. Harutyunyan[1], and A.M. Ishkhanyan[1,3]

[1]Institute for Physical Research, NAS of Armenia, Ashtarak, 0203 Armenia
[2]Moscow Institute of Physics and Technology, Dolgoprudny, 141700 Russia
[3]Institute of Physics and Technology, National Research Tomsk Polytechnic University, Tomsk, 634050 Russian Federation

We introduce a new exactly integrable potential for the Schrödinger equation for which the solution of the problem may be expressed in terms of the Gauss hypergeometric functions. This is a potential step with variable height and steepness. We present the general solution of the problem, discuss the transmission of a quantum particle above the barrier, and derive explicit expressions for the reflection and transmission coefficients.

**PACS numbers:** 03.65.Ge Solutions of wave equations: bound states, 02.30.Ik Integrable systems, 02.30.Gp Special functions

**Keywords:** Schrödinger equation, solvable potential, Gauss hypergeometric function

## 1. Introduction

The list of the solutions of the one-dimensional stationary Schrödinger equation in terms of the Gauss ordinary hypergeometric functions for potentials for which all the involved parameters are varied independently includes just three names. These are the Eckart [1] and Pöschl-Teller [2] classical potentials known from the early days of quantum mechanics and the third hypergeometric potential introduced recently [3]. In this paper we introduce another potential belonging to this class.

This is an asymmetric potential-step with variable height and steepness. Though independent, the potential has much in common with the third exactly integrable hypergeometric potential which is also an asymmetric step-barrier [3]. In particular, both potentials are four-parameter and belong to the same general Heun family [4,5]. Besides, in both cases, the general solution of the problem has been expressed in terms of fundamental solutions, each of which is represented by an irreducible linear combination of two ordinary hypergeometric functions. It has also been shown that this combination may be represented by a single generalized hypergeometric function [6,7].

To derive the solution of the Schrödinger equation, we have applied a recently proposed approach suitable for the identification of integrable potentials that are proportional to an energy-independent parameter and have a shape that does not depend on that parameter [8]. The approach may also be used to generate conditionally integrable potentials (see, e.g., several classical [9-14] and recent [15-18] examples), as well as quasi-exactly solvable potentials (see, e.g., [19-21]).

In this paper we present the explicit solution and discuss the quantum mechanical reflection and transmission of a quantum particle above the barrier. The reflection and transmission coefficients are given by simple formulas.

## 2. The potential and reduction to the Heun equation

The potential we introduce is given as

$$V = V_0 + \frac{V_1}{z} \qquad (1)$$

with

$$z = -1 + \frac{1}{\left(e^{x/(2\sigma)} + \sqrt{1+e^{x/\sigma}}\right)^{2/3}} + \left(e^{x/(2\sigma)} + \sqrt{1+e^{x/\sigma}}\right)^{2/3}, \qquad (2)$$

where one may replace $x$ by $x-x_0$ with arbitrary $x_0$. The shape of the potential is shown in Fig. 1. This is an asymmetric step-barrier with controllable height and steepness. The potential involves four independent parameters, $V_0, V_1, \sigma$ and $x_0$, which stand, respectively, for the energy origin, the height, the steepness and the space position of the step (obviously, without any loss of the generality, one may put $V_0 = 0$ and $x_0 = 0$).

To solve the one-dimensional stationary Schrödinger equation

$$\frac{d^2\psi}{dx^2} + \frac{2m}{\hbar^2}(E - V(x))\psi = 0 \qquad (3)$$

for the potential (1),(2), we reduce it to the general Heun equation [22-25]

$$\frac{d^2u}{dz^2} + \left(\frac{\gamma}{z-a_1} + \frac{\delta}{z-a_2} + \frac{\varepsilon}{z-a_3}\right)\frac{du}{dz} + \frac{\alpha\beta z - q}{(z-a_1)(z-a_2)(z-a_3)} u = 0. \qquad (4)$$

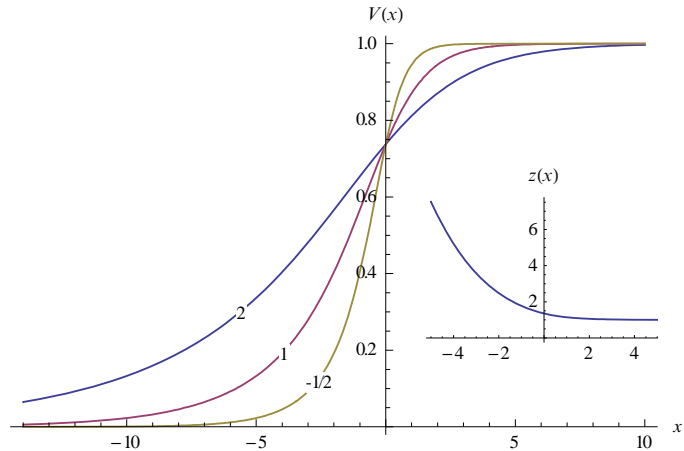

Fig.1. Potential (1) for $V_0 = 0$, $V_1 = 1$, $x_0 = 0$, $\sigma = -1/2, -1, -2$ (in units $\hbar = m = 1$). The inset presents the coordinate transformation $z(x)$ for $\sigma = -1$.

The Heun equation is a natural generalization of the Gauss hypergeometric equation. It is the most general Fuchsian second-order linear ordinary differential equation possessing four regular singularities, i.e., one more than those of the Gauss equation. Note that equation (4) assumes the finite singularities being located at the points $z = a_1, a_2, a_3$.

The reduction of the Schrödinger equation to the general Heun equation is achieved by the variable change $\psi = \varphi(z) u(z)$, $z = z(x)$, which results in the following equation for the new unknown function $u(z)$:

$$u_{zz} + \left(2\frac{\varphi_z}{\varphi} + \frac{\rho_z}{\rho}\right) u_z + \left(\frac{\varphi_{zz}}{\varphi} + \frac{\rho_z}{\rho}\frac{\varphi_z}{\varphi} + \frac{2m}{\hbar^2}\frac{E-V(z)}{\rho^2}\right) u = 0, \tag{5}$$

where $\rho = dz/dx$. Consider the coordinate transformation given by equation (2). We note that the function $z(x)$ is the real root of the cubic equation

$$(z+2)^2 (z-1) = 4 e^{x/\sigma}, \tag{6}$$

from which we obtain

$$\rho = \frac{dz}{dx} = \frac{(z+2)(z-1)}{3\sigma z}. \tag{7}$$

With this, it is readily checked that the pre-factor

$$\varphi(z) = (z+2)^{\alpha_1} (z-1)^{\alpha_2} \tag{8}$$

reduces equation (5) to the Heun equation with singularities $(a_1, a_2, a_3) = (-2, 1, 0)$.

$$\frac{d^2 u}{dz^2} + \left(\frac{\gamma}{z+2} + \frac{\delta}{z-1} + \frac{\varepsilon}{z}\right) \frac{du}{dz} + \frac{\alpha\beta z - q}{(z+2)(z-1)z} u = 0 \tag{9}$$

with the involved parameters given as

$$(\gamma, \delta, \varepsilon) = (1 + 2\alpha_1, 1 + 2\alpha_2, -1), \tag{10}$$

$$(\alpha\beta, q) = \left((\alpha_1 + \alpha_2)^2 + \frac{18 m \sigma^2 (E - V_0)}{\hbar^2}, -\alpha_1 + 2\alpha_2\right), \tag{11}$$

and the exponents $\alpha_1, \alpha_2$ of the pre-factor (8) being

$$(\alpha_1, \alpha_2) = \left(\pm 2i\sigma \sqrt{\frac{2m}{\hbar^2}\left(E - V_0 + \frac{V_1}{2}\right)}, \pm i\sigma \sqrt{\frac{2m}{\hbar^2}(E - V_0 - V_1)}\right). \tag{12}$$

We note that any combination of plus and minus signs for $\alpha_1, \alpha_2$ is applicable. Hence, by choosing different combinations, one can derive different fundamental solutions.

## 3. The solution of the Heun equation

The next step is to construct the solution of the Heun equation (9) in terms of the hypergeometric functions. One first checks that the known direct Heun-to-hypergeometric reductions (see, e.g., [26-30]) do not help. Instead, one may apply the series solutions in terms of the Gauss hypergeometric functions [31-35]. The cases when such expansions terminate thus resulting in finite-sum closed-form solutions have been discussed in [36]. It was shown that the series terminates if a characteristic exponent of a singularity of the Heun equation is a positive integer and the accessory parameter $q$ satisfies a polynomial equation.

In our case $\varepsilon = -1$ so that a characteristic exponent of the singularity $z = 0$ is $\mu = 1 - \varepsilon = 2$. Hence, a solution of the Heun equation in the form of a finite-sum of the Gauss hypergeometric functions may exist. For this to be the case, the necessary condition is

$$q^2 + q(1+\gamma-2\delta) - 2\alpha\beta = 0. \tag{13}$$

We present the detailed derivation of this equation in Appendix. It may be readily checked that the equation is satisfied by parameters (10)-(12). Accordingly, a fundamental solution of the Heun equation reads (see the Appendix)

$$u_1 = {}_2F_1\left(\alpha,\beta;\gamma;\frac{z+2}{3}\right) + \frac{\gamma-1}{q-2(\delta-1)} \; {}_2F_1\left(\alpha,\beta;\gamma-1;\frac{z+2}{3}\right). \tag{14}$$

In the similar way, the second independent fundamental solution is shown to be

$$u_2 = {}_2F_1\left(\alpha,\beta;\delta;\frac{1-z}{3}\right) + \frac{q(1-\delta)}{\alpha\beta-q(1-\delta)} \; {}_2F_1\left(\alpha,\beta;\delta-1;\frac{1-z}{3}\right). \tag{15}$$

Thus, the general solution of the Schrödinger equation may be written as

$$\psi(x) = (z+2)^{\alpha_1}(z-1)^{\alpha_2}(c_1 u_1 + c_2 u_2) \tag{16}$$

with arbitrary $c_1$, $c_2$. This solution is valid for arbitrary (real or complex) set of all involved parameters (with the proviso that none of $\gamma$ and $\delta$ is zero, one, or a negative integer).

We conclude this section by noting that Letessier noticed [37] (see also [38]) that the fundamental solutions (14),(15) can be represented by the Clausen generalized hypergeometric functions [39]:

$$u_1 = {}_3F_2\left(\alpha,\beta,1+\frac{2\alpha\beta}{q};\frac{2\alpha\beta}{q},\gamma;\frac{z+2}{3}\right), \quad u_2 = {}_3F_2\left(\alpha,\beta,1-\frac{\alpha\beta}{q};-\frac{\alpha\beta}{q},\delta;\frac{1-z}{3}\right). \tag{17}$$

This is a useful form that in several cases provides simpler derivations.

## 4. Above-barrier transmission

To discuss the quantum-mechanical reflection on transmission of a particle above the potential barrier, that is, when $E > V_0 + V_1$, we note that the coordinate transformation given by equation (2) maps the real axes $x \in (-\infty, +\infty)$ onto the interval $z \in (1, +\infty)$ and that the asymptotes at infinity for $\sigma < 0$ are

$$z|_{x \to -\infty} \sim 2^{2/3} e^{x/(3\sigma)} \to +\infty, \quad z|_{x \to +\infty} \sim 1 + 4 e^{x/\sigma}/9 \to 1. \tag{18}$$

Let the exponents $\alpha_1$ and $\alpha_2$ take the plus sign in equations (12). Then, the pre-factor of the general solution (16) at $x \to +\infty$ behaves as

$$\varphi = (z+2)^{\alpha_1}(z-1)^{\alpha_2} \sim 2^{2\alpha_2} 3^{\alpha_1 - 2\alpha_2} e^{ik_2 x}, \quad k_2 = \sqrt{\frac{2m}{\hbar^2}(E - V_0 - V_1)}. \tag{19}$$

Demanding now that the wave function at $x \to +\infty$ involves only one plane wave (that is, only the transmitted wave), we derive $c_1 = 0$ and

$$\psi(+\infty) \sim C e^{ik_2 x}, \quad C = \frac{2^{2\alpha_2} 3^{1+\alpha_1 - 2\alpha_2}}{\alpha_2(-\alpha_1 + 2\alpha_2)} \frac{m\sigma^2 V_1}{\hbar^2} c_2. \tag{20}$$

Expanding then the solution at $x \to -\infty$, we arrive at the asymptote

$$\psi(-\infty) \sim A e^{+ik_1 x} + B e^{-ik_1 x}, \quad k_1 = \sqrt{\frac{2m}{\hbar^2}(E - V_0)} \tag{21}$$

with

$$A = -\frac{2^{2ik_1\sigma} 3^{1+\alpha_1+\alpha_2-3ik_1\sigma}(\alpha_2 - ik_1\sigma)\Gamma(2\alpha_2)\Gamma(6i\sigma k_1)}{(\alpha_1 - 2\alpha_2)\Gamma(3i\sigma k_1 - \alpha_1 + \alpha_2)\Gamma(3i\sigma k_1 + \alpha_1 + \alpha_2)} c_2, \tag{22}$$

$$B = -\frac{2^{-2ik_1\sigma} 3^{1+\alpha_1+\alpha_2+3ik_1\sigma}(\alpha_2 + ik_1\sigma)\Gamma(2\alpha_2)\Gamma(-6i\sigma k_1)}{(\alpha_1 - 2\alpha_2)\Gamma(-3i\sigma k_1 - \alpha_1 + \alpha_2)\Gamma(-3i\sigma k_1 + \alpha_1 + \alpha_2)} c_2, \tag{23}$$

where $\Gamma$ is the Euler gamma-function. The transmission coefficient is then given as

$$T = \left|\frac{k_2 C^2}{k_1 A^2}\right| = \frac{\sinh(6\pi\sigma k_1)\sinh(2\pi\sigma k_2)}{\sinh[\pi\sigma(3k_1 + k_2 - 2k_h)]\sinh[\pi\sigma(3k_1 + k_2 + 2k_h)]}, \tag{24}$$

where we have introduced the notation

$$k_h = \sqrt{\frac{2m}{\hbar^2}\left(E - V_0 + \frac{V_1}{2}\right)}. \tag{25}$$

In the limit $\sigma \to 0$ we have

$$T = \frac{4 k_1 k_2}{(k_1 + k_2)^2}, \tag{26}$$

which is the result for the abrupt-step potential [40]. As expected, the correction term is shown to be positive, hence, because of the smoothness, the transmission above the potential (1),(2) is always greater than that for the abrupt-step potential. In the idealized infinitely-smooth limit $\sigma \to \infty$ the potential becomes transparent. The reflection coefficient $R = 1 - T$ is shown in Fig. 2.

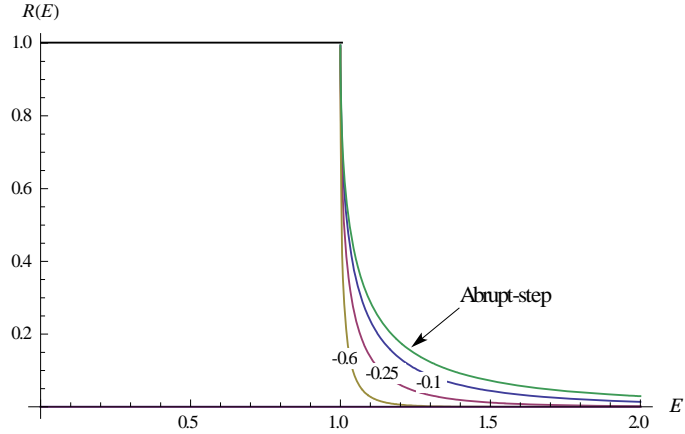

Fig.2. The reflection coefficient $R = 1 - T$ versus energy $E$ for $(V_0, V_1, x_0, m, h) = (0, 1, 0, 1, 1)$, $\sigma = -0.1, -0.25, -0.6$.

## 5. Discussion

Thus, we have introduced one more quantum mechanical potential for which the one-dimensional stationary Schrödinger equation is solved in terms of the Gauss hypergeometric functions [1-3]. The new potential is a step-barrier with controllable height and width. We have discussed the transmission of a quantum particle above this barrier.

The potential is a four-parameter sub-potential of the fifth eight-parameter general Heun family defined by the triad $(m_1, m_2, m_3) = (1, 1, -1)$ [25]. This is a remarkable family in that it generalizes the Eckart potential and involves, as independent particular cases, the third exactly integrable hypergeometric potential and the potential we have introduced. In addition, the family contains a number of conditionally integrable sub-potentials, in particular, the Dutt-Khare-Varshni [12] and López-Ortega potentials [16], as well as a variety of quasi-exactly solvable potentials [21] or fixed-energy solutions [41].

An observation worth of some attention is that the two new hypergeometric potentials (as well as other recently reported exactly integrable potentials of the Heun class [42-44]) are four-parameter, while the classical ordinary hypergeometric potentials by Eckart and Pöschl-Teller (as well as the three classical confluent hypergeometric potentials [45-48]) are five-parameter. We wonder if there exists a more general exactly integrable hypergeometric reduction of the fifth general Heun family, which involves as particular cases the two recent hypergeometric potentials. We hope to explore this possibility in the near future.

**Acknowledgments**

This research has been conducted within the scope of the International Associated Laboratory IRMAS (CNRS-France & SCS-Armenia). The work has been supported by the Armenian State Committee of Science (SCS Grant No. 18RF-139), Armenian National Science and Education Fund (ANSEF Grant No. PS-4986) and the project "Leading Russian Research Universities" (Grant No. FTI_24_2016 of the Tomsk Polytechnic University). T.A. Ishkhanyan acknowledges the support from SPIE through a 2017 Optics and Photonics Education Scholarship, and thanks the French Embassy in Armenia for a doctoral grant as well as the Agence universitaire de la Francophonie for a Scientific Mobility grant.

**Appendix**

Consider the general Heun equation written in its canonical form [23-25]:

$$\frac{d^2u}{dy^2} + \left(\frac{\gamma}{y} + \frac{\delta}{y-1} + \frac{\varepsilon}{y-a}\right)\frac{du}{dy} + \frac{\alpha\beta y - q_0}{y(y-1)(y-a)}u = 0, \quad (A1)$$

where the parameters obey the Fuchsian condition

$$1 + \alpha + \beta = \gamma + \delta + \varepsilon. \quad (A2)$$

We examine an expansion of the solution of equation (A1) of the form

$$u = \sum_{n=0}^{\infty} c_n \cdot {}_2F_1(\alpha, \beta; \gamma - n; y) = \sum_{n=0}^{\infty} c_n w_n. \quad (A3)$$

The involved Gauss hypergeometric functions $w_n$ obey the equation

$$\frac{d^2 w_n}{dy^2} + \left(\frac{\gamma - n}{y} + \frac{\delta_n}{y-1}\right)\frac{dw_n}{dy} + \frac{\alpha\beta}{y(y-1)}w_n = 0, \quad (A4)$$

where (compare with (A2))

$$1 + \alpha + \beta = \gamma - n + \delta_n. \quad (A5)$$

In order to match the last equation with the Fuchsian condition (A2) for all $n$, we put $\delta_n = \delta + \varepsilon + n$. Then, substituting equations (A3),(A4) into equation (A1), we derive

$$\sum_n c_n \left(\left(\frac{n}{y} - \frac{\varepsilon + n}{y-1} + \frac{\varepsilon}{y-a}\right)\frac{dw_n}{dy} + \frac{\alpha\beta a - q_0}{y(y-1)(y-a)}w_n\right) = 0 \quad (A6)$$

or

$$\sum_n c_n \left((an + (a\varepsilon - \varepsilon - n)y)\frac{dw_n}{dy} + (\alpha\beta a - q_0)w_n\right) = 0. \quad (A7)$$

Now, using the following relations between the involved hypergeometric functions [24]:

$$\frac{dw_n}{dy} = (\gamma - n - 1)w_{n+1} + (1 + \alpha + \beta + 2n - 2\gamma)w_n + \frac{(\alpha + n - \gamma)(\beta + n - \gamma)}{\gamma - n}w_{n-1}, \quad (A8)$$

$$y\frac{dw_n}{dy} = (\gamma - n - 1)(w_{n+1} - w_n), \quad (A9)$$

equation (A7) is rewritten as

$$\sum_n c_n \left( \frac{an(\alpha + n - \gamma)(\beta + n - \gamma)}{\gamma - n}w_{n-1} + (a-1)(\varepsilon + n)(\gamma - n - 1)w_{n+1} + \left(-(a-1)(\varepsilon + n)(\gamma - n - 1) + an(\alpha + \beta - \gamma + n) + a\alpha\beta - q_0\right)w_n \right) = 0, \quad (A10)$$

from which we obtain a three-term recurrence relation for the coefficients of the expansion:

$$R_n c_n + Q_{n-1} c_{n-1} + P_{n-2} c_{n-2} = 0 \quad (A11)$$

with

$$R_n = \frac{an}{\gamma - n}(\alpha - \gamma + n)(\beta - \gamma + n), \quad (A12)$$

$$Q_n = -P_n + an(\alpha + \beta - \gamma + n) + a\alpha\beta - q_0, \quad (A13)$$

$$P_n = (a-1)(\varepsilon + n)(\gamma - n - 1). \quad (A14)$$

The constructed expansion terminates if $c_N \neq 0$ and $c_{N+1} = c_{N+2} = 0$ for some $N = 0, 1, 2, \ldots$. The recurrence relation (A11) for $n = N + 2$ is then reduced to $P_N = 0$, which for $a \neq 1$ and non-integer $\gamma$ is satisfied if

$$\varepsilon = -N, \quad (A15)$$

and the equation $c_{N+1} = 0$ leads to a polynomial equation of degree $N + 1$ for the accessory parameter $q$. For $\varepsilon = -1$ this equation reads

$$q_0^2 + q_0(\gamma - 1 - a(\alpha + \beta) - 2a\alpha\beta) + a\alpha\beta(a(1 + \alpha + \beta) - \gamma + a\alpha\beta) = 0. \quad (A16)$$

The resulting solution of the Heun equation may be readily derived to be

$$u = {}_2F_1(\alpha, \beta; \gamma; y) + \frac{(\gamma - 1)(q + \gamma - 1)}{2(1 + \alpha - \gamma)(1 + \beta - \gamma)} {}_2F_1(\alpha, \beta; \gamma - 1; y) \quad (A17)$$

Let us now return to the general Heun equation (9). By the linear change $z = 3y - 2$ this equation is rewritten as

$$\frac{d^2u}{dy^2} + \left(\frac{\gamma}{y} + \frac{\delta}{y-1} + \frac{\varepsilon}{y - 2/3}\right)\frac{du}{dy} + \frac{\alpha\beta y - (q + 2\alpha\beta)/3}{y(y-1)(y-2/3)}u = 0. \quad (A18)$$

Comparing with equation (A1), we have

$$a = \frac{2}{3}, \quad q_0 = \frac{q + 2\alpha\beta}{3}. \tag{A19}$$

With this, equation (A16) reproduces equation (13):

$$q^2 + q(1 + \gamma - 2\delta) - 2\alpha\beta = 0, \tag{A20}$$

and equation (A17) with $y = (z+2)/3$ gives the solution (14):

$$u = {}_2F_1\left(\alpha, \beta; \gamma; \frac{z+2}{3}\right) + \frac{\gamma - 1}{q - 2(\delta - 1)} {}_2F_1\left(\alpha, \beta; \gamma - 1; \frac{z+2}{3}\right). \tag{A21}$$

**References**

1. C. Eckart, "The penetration of a potential barrier by electrons", Phys. Rev. **35**, 1303-1309 (1930).
2. G. Pöschl, E. Teller, "Bemerkungen zur Quantenmechanik des anharmonischen Oszillators", Z. Phys. **83**, 143-151 (1933).
3. A.M. Ishkhanyan, "The third exactly solvable hypergeometric quantum-mechanical potential", EPL **115**, 20002 (2016).
4. A.M. Ishkhanyan, "Schrödinger potentials solvable in terms of the general Heun functions", Ann. Phys. **388**, 456-471 (2018).
5. A. Lemieux and A.K. Bose, "Construction de potentiels pour lesquels l'équation de Schrödinger est soluble", Ann. Inst. Henri Poincaré A **10**, 259-270 (1969).
6. W.N. Bailey, Generalized Hypergeometric Series (Stechert-Hafner Service Agency, New York and London 1964).
7. L.J. Slater, *Generalized hypergeometric functions* (Cambridge University Press, Cambridge, 1966).
8. A. Ishkhanyan and V. Krainov, "Discretization of Natanzon potentials", Eur. Phys. J. Plus **131**, 342 (2016).
9. F.H. Stillinger, "Solution of a quantum mechanical eigenvalue problem with long range potentials", J. Math. Phys. **20**, 1891-1895 (1979).
10. J.N. Ginocchio, "A class of exactly solvable potentials. I. One-dimensional Schrödinger equation", Ann. Phys. **152**, 203-219 (1984).
11. H. Exton, "The exact solution of two new types of Schrodinger equation", J. Phys. A **28**, 6739-6741 (1995).
12. R. Dutt, A. Khare, and Y.P. Varshni, "New class of conditionally exactly solvable potentials in quantum mechanics", J. Phys. A **28**, L107-L113 (1995).
13. C. Grosche, "Conditionally solvable path integral problems", J. Phys. A **28**, 5889-5902 (1995).
14. B.W. Williams, "Exact solutions of a Schrödinger equation based on the Lambert function", Phys. Lett. A **334**, 117-122 (2005).
15. A. López-Ortega, "New conditionally exactly solvable inverse power law potentials", Phys. Scr. **90**, 085202 (2015).
16. A. López-Ortega, "A conditionally exactly solvable generalization of the potential step", arXiv:1512.04196 [math-ph] (2015).
17. A. López-Ortega, "New conditionally exactly solvable potentials of exponential type", arXiv:1602.00405 [math-ph] (2016).

18. T.A. Ishkhanyan and A.M. Ishkhanyan, "Solutions of the bi-confluent Heun equation in terms of the Hermite functions", Ann. Phys. **383**, 79-91 (2017).
19. A.G. Ushveridze, *Quasi-exactly solvable models in quantum mechanics* (IOP Publishing, Bristol, 1994).
20. C.M. Bender and M. Monou, "New quasi-exactly solvable sextic polynomial potentials", JPA **38**, 2179-2187 (2005).
21. A.V. Turbiner, "One-dimensional quasi-exactly solvable Schrödinger equations", Physics Reports **642**, 1-71 (2016).
22. K. Heun, "Zur Theorie der Riemann'schen Functionen Zweiter Ordnung mit Verzweigungspunkten", Math. Ann. **33**, 161 (1889).
23. A. Ronveaux (ed.), *Heun's Differential Equations* (Oxford University Press, London, 1995).
24. S.Yu. Slavyanov and W. Lay, *Special functions* (Oxford University Press, Oxford, 2000).
25. F.W.J. Olver, D.W. Lozier, R.F. Boisvert, and C.W. Clark (eds.), *NIST Handbook of Mathematical Functions* (Cambridge University Press, New York, 2010).
26. K. Kuiken, "Heun's equation and the hypergeometric equation", SIAM J. Math. Anal. **10**, 655-657 (1979).
27. R.S. Maier, "On reducing the Heun equation to the hypergeometric equation", J. Diff. Equations **213**, 171 (2005).
28. R. Vidunas, G. Filipuk, "Parametric transformations between the Heun and Gauss hypergeometric functions", Funkcialaj Ekvacioj **56**, 271-321 (2013).
29. R. Vidunas, G. Filipuk, "A classification of coverings yielding Heun-to-hypergeometric reductions", Osaka J. Math. **51**, 867-903 (2014).
30. M. van Hoeij and R. Vidunas, "Belyi coverings for hyperbolic Heun-to-hypergeometric transformations", J. Algebra. **441**, 609-659 (2015).
31. N. Svartholm, "Die Lösung der Fuchs'schen Differentialgleichung zweiter Ordnung durch Hypergeometrische Polynome", Math. Ann. **116**, 413-421 (1939).
32. A. Erdélyi, "The Fuchsian equation of second order with four singularities", Duke Math. J. **9**, 48-58 (1942).
33. A. Erdélyi, "Certain expansions of solutions of the Heun equation", Q. J. Math. (Oxford) **15**, 62-69 (1944).
34. D. Schmidt, "Die Lösung der linearen Differentialgleichung 2. Ordnung um zwei einfache Singularitäten durch Reihen nach hypergeometrischen Funktionen", J. Reine Angew. Math. **309**, 127-148 (1979).
35. E.G. Kalnins and W. Miller, Jr., "Hypergeometric expansions of Heun polynomials", SIAM J. Math. Anal. **22**, 1450-1459 (1991).
36. A. Hautot, "Sur des combinaisons linéaires d'un nombre fini de fonctions transcendantes comme solutions d'équations différentielles du second ordre", Bull. Soc. Roy. Sci. Liège **40**, 13-23 (1971).
37. J. Letessier, "Co-recursive associated Jacobi polynomials", J. Comp. Appl. Math. **57**, 203-213 (1995).
38. R.S. Maier, "P-symbols, Heun identities, and $_3F_2$ identities", Contemporary Mathematics **471**, 139-159 (2008).
39. T. Clausen, ""Ueber die Fälle, wenn die Reihe von der Form ... ein Quadrat von der Form ... hat",", J. Reine Angew. Math. **3**, 89-91 (1828).
40. S. Flügge, *Practical Quantum Mechanics I, II* (Springer Verlag, Berlin, 1971).
41. B.-H. Chen, Y. Wu and Q.-T. Xie, "Heun functions and quasi-exactly solvable double-well potentials", J. Phys. A **46**, 035301 (2013).
42. A.M. Ishkhanyan, "Exact solution of the Schrödinger equation for the inverse square root

potential $V_0/\sqrt{x}$ ", EPL **112**, 10006 (2015).
43. A.M. Ishkhanyan, "The Lambert-W step-potential - an exactly solvable confluent hypergeometric potential", Phys. Lett. A **380**, 640-644 (2016).
44. A.M. Ishkhanyan, "A singular Lambert-W Schrödinger potential exactly solvable in terms of the confluent hypergeometric functions", Mod. Phys. Lett. A **31**, 1650177 (2016).
45. A. Kratzer, "Die ultraroten Rotationsspektren der Halogenwasserstoffe", Z. Phys. **3**, 289-307 (1920).
46. E. Schrödinger, "Quantisierung als Eigenwertproblem (Erste Mitteilung)", Annalen der Physik **76**, 361-376 (1926).
47. E. Schrödinger, "Quantisierung als Eigenwertproblem (Zweite Mitteilung)". Annalen der Physik **79**, 489-527 (1926).
48. P.M. Morse, "Diatomic molecules according to the wave mechanics. II. Vibrational levels", Phys. Rev. **34**, 57-64 (1929).